\documentclass[letterpaper]{article}%
\usepackage{amsmath}%
\usepackage{amsfonts}%
\usepackage{amssymb}%
\usepackage{graphicx}
\usepackage[draft]{hyperref}
\usepackage[margin=2.5cm]{geometry}
\usepackage[rgb]{xcolor}
\usepackage{ifpdf}

\begin{document}

\title{\textbf{Nanobeam Photonic Crystal Cavity Light-Emitting Diodes}}
\author{Gary Shambat,$^{1,*}$ Bryan Ellis,$^1$ Jan Petykiewicz,$^1$ 
\\Marie A. Mayer,$^2$ Tomas Sarmiento,$^1$ James Harris,$^1$ Eugene E. Haller,$^2$ 
\\and Jelena Vu\v{c}kovi\'{c}$^1$
\\ \\{\small $^1$Department of Electrical Engineering, Stanford University, Stanford, California 94305 USA}
\\{\small$^2$Materials Sciences Division, Lawrence Berkeley National Laboratory, Berkeley, California 94720}  \\{\small$^*$email: gshambat@stanford.edu}}

\date{}
\maketitle

\begin{abstract}
We present results on electrically driven nanobeam photonic crystal cavities formed out of a lateral \textit{p-i-n} junction in gallium arsenide. Despite their small conducting dimensions, nanobeams have robust electrical properties with high current densities possible at low drive powers. Much like their two-dimensional counterparts, the nanobeam cavities exhibit bright electroluminescence at room temperature from embedded 1,250 nm InAs quantum dots. A small room temperature differential gain is observed in the cavities with minor beam self-heating suggesting that lasing is possible. These results open the door for efficient electrical control of active nanobeam cavities for diverse nanophotonic applications.  
\end{abstract}

\section{Main Text}

Photonic crystal (PC) cavities have attracted much attention in the last decade for their strong light-matter interaction properties. Recently, there has been significant interest in one-dimensional nanobeam PC cavities due to their smaller footprint and exceptional quality (Q) -factors, exceeding those of their two-dimensional counterparts \cite{notomi}. In addition, 1D nanobeams are well suited for coupling light to on-chip networks \cite{quan}, as well as for MEMS manipulation tuning \cite{frank}. Researchers have been able to exploit these properties to develop low threshold lasers \cite{zhang, yiyang}, optomechanic crystals \cite{matt}, strongly coupled quantum dot (QD) cavity systems \cite{ohta}, as well as chemical sensors \cite{wang}. 

However, there is a need for efficient electrical injection in active nanobeam devices, which has not yet been addressed. Recently, we demonstrated a world record low threshold PC laser using a lateral \textit{p-i-n} junction formed by ion implantation \cite{ellis}. This type of doping layout allows for precise current flow that can be defined lithographically and be applied to arbitrary cavity designs. The previous laser diode was formed out of an L3 PC cavity defect \cite{akahane} embedded in a 2D membrane and exhibited lasing with a 208 nW threshold at 50K. In this letter, we show that the same lateral junction process can be applied to 1D nanobeam photonic crystal cavities, yielding room temperature light-emitting diodes (LEDs) at low control powers. In spite of their narrow cross sections, nanobeams can nonetheless direct current efficiently through the sidewalls and deliver charge to the central cavities. The results presented here are a promising step toward practical active nanobeam device architectures.       

Electrically contacted nanobeams were fabricated using a similar process flow as in \cite{ellis}. We start with a 220 nm thick GaAs membrane with three layers of embedded high density (300 dots/$\mu$m$^2$) InAs quantum dots. Electron beam lithography steps were used to pattern windows in a deposited silicon nitride mask to implant N- (silicon) and P- (beryllium) type dopants. A rapid thermal anneal activated the N and P regions to 6x10$^{17}$ and 2.5x10$^{19}$ cm$^{-3}$ doping densities, respectively, and also blueshifted the peak of the QD ensemble emission from 1,310 nm to 1,230 nm. E-beam lithography and dry etching next defined the photonic crystal cavity patterns along with isolation trenches in the GaAs membrane. We chose a common nanobeam cavity design that incorporates a 5-hole linear taper in lattice constant (from a = 322 to a = 266) and hole radius (r = 0.22a) to increase the quality factor \cite{deotare}. The beam widths were 500-600 nm across. Metal contacts were then deposited and the beams were released by removing the sacrificial AlGaAs layer. Figure 1(a-c) shows several scanning electron microscope (SEM) close-up images of a fabricated nanobeam. From the tilted image in Figure 1(b), it is clear that the beams exhibit a small bowing effect likely due to strain from the underlying AlGaAs layer. We do not believe this sagging effect has much of an impact on device performance because of the similarities in results between beams and non-bowing 2D structures. Figure 1(d) shows a finite-difference time domain (FDTD) simulation of the electric field of the cavity design, with a theoretical quality (Q) -factor of 95,000.    

Electrical measurements were performed at room temperature using a Keithley 2635 sourcemeter with sub-nanoamp resolution and standard electrical probes. Figure 2(a) shows the measured IV characteristics for beams with intrinsic region widths of 400 nm and 5 $\mu$m as well as for a control sample beam with no holes and an intrinsic region of 400 nm. As expected the current is substantially reduced by two orders of magnitude when the intrinsic region is large due to the poor diffusion of carriers across the junction. Interestingly, the current magnitudes are very similar for both control and regular samples, with a current near 10 $\mu$A for a 1.2 V bias. Despite the reduced current cross sectional area of the beams with holes, the presence of additional etched surfaces creates a greater recombination current. Therefore the narrow 100-200 nm conducting sidewalls of the beams are large enough for efficient carrier flow. The series resistance of 9,000 $\Omega$ agrees well with the geometrical sheet resistance of our structure and is correspondingly larger than the 1,150 $\Omega$ measured in 2D membranes \cite{ellis}. We perform 2D Poisson simulations of our devices using the Sentarus package by incorporating measured doping densities, mobilities, and non-radiative recombination lifetime values \cite{shambat}. Full details of our method will be reported elsewhere. Figure 2(a) displays the simulation current data alongside the experimental data, where reasonable agreement is seen for all three structures. The slightly lower simulation values are likely due to the exclusion of top and bottom device surfaces, for which extra recombination current would be observed. Also, the experimental intrinsic region width is slightly narrower than the simulated width due to dopant diffusion, giving rise to greater current. In Figure 2(b), we see the steady state carrier densities for a 1.2 V bias along the beam length with a 150 nm offset from the center axis. The injected electron and hole densities are only 4x10$^{15}$ cm$^{-3}$ due to the fast (6 ps) non-radiative recombination lifetime. Two-dimensional plots of the carrier densities for a 1.2 V bias are seen in Figures 2(c-d), where it is clear that the minority carriers are well localized to the cavity region. Finally, Figure 2(e) shows a map of the current density at 1.2 V for which high currents are observed in the beam sidewalls at values up to 10 kA/cm$^2$.  

Electroluminescence (EL) data was taken at room temperature by forward biasing the nanobeam diodes and collecting the emission with a spectrometer and liquid nitrogen cooled InGaAs CCD array detector. Figure 3(a) displays the output spectrum for a nanobeam biased to 5 $\mu$A. Bright cavity mode emission is seen superimposed upon the weaker QD background, indicating successful carrier injection into the nanobeam cavity. The cavity mode has a quality factor of 2,900 (inset in Figure 3(a)), well below the theoretical value of 95,000. We believe the quality factor is limited both by fabrication imperfections as well as free carrier absorption by the nearby doping regions. As seen in Figure 3(c), the IR output emission is heavily concentrated to the nanobeam cavity center with a small amount of scattered emission visible at the nanobeam edges. We next investigate the properties of the nanobeams as we increase the injection current. For this experiment, a different cavity than the one seen in Figure 3(a) was used due to accidental device failure. The spectrum of the new cavity is seen in Figure 3(b) where the mode Q-factor is reduced to about 500. In Figure 3(d), the cavity output emission and Q-factor are plotted versus injection current. The cavity power output is linear for the entire range and therefore no lasing is observed. A small amount of room temperature linewidth narrowing was observed as the Q-factor increased with injection current. Therefore, we believe it is possible to obtain lasing in these important nanophotonic structures with much higher Q cavities \cite{nomura}.

In order to characterize the heating effects of electrically pumped nanobeams, we examine the mode peak wavelength as a function of injection current. Figure 4(a) shows that the mode wavelength shifts by less than 1 nm for 5 $\mu$A of injection current. The thermal dependence of the refractive index of GaAs near 1.3 $\mu$m is given by \textit{dn} = 2.7x10$^{-4}$\textit{dT} K$^{-1}$, where \textit{dT} is the change in temperature and \textit{dn} is the change in material refractive index \cite{talghader}. Hence the cavity wavelength is expected to shift via second order perturbation theory as \textit{d$\lambda$/$\lambda$} = \textit{dn/n}, where \textit{d$\lambda$} is the change in mode wavelength, \textit{$\lambda$} is the cavity peak wavelength, and \textit{n} is the nominal material refractive index. For a measured \textit{d$\lambda$} of 0.6 nm, \textit{$\lambda$} = 1,266 nm, and n = 3.5, the calculated temperature rise is only 6K. For comparison, we calculate the lattice temperature from a hydrodynamic transport model in Sentaurus and find that the heating is ~3.3K at the cavity center. Therefore, our electrical design is very robust against self-heating despite the large injection current densities.    

In summary, we have demonstrated efficient electrically driven photonic crystal nanobeam cavity LEDs at room temperature. The results here are an extension of our lateral \textit{p-i-n} junction design in 2D PCs showing the versatility of the fabrication technique. Our nanobeams have excellent electrical properties even with their narrow conducting paths. Future designs could incorporate surface passivation techniques to slow down non-radiative recombination and hence increase the charge injection level. Other geometrical modifications such as beam width and hole size could further optimize the device performance. The electrical control of nanobeam cavities demonstrated here is an important step forward in developing practical on-chip devices for diverse applications such as lasers, sensors, and optomechanics.  

\subsection{Acknowledgements}
Gary Shambat and Bryan Ellis were supported by the Stanford Graduate Fellowship. Gary
Shambat is also supported by the NSF GRFP. The authors acknowledge the support of the
Interconnect Focus Center, the AFOSR MURI for Complex and Robust On-chip Nanophotonics
(Dr. Gernot Pomrenke), grant number FA9550-09-1-0704, and the Director, Office of Science,
Office of Basic Energy Sciences, Materials Sciences and Engineering Division, of the US
Department of Energy under Contract No. DE-AC02-05CH11231. Work was performed in part at the Stanford Nanofabrication Facility of NNIN supported by the National Science Foundation. We also acknowledge Kelley Rivoire for assisting in SEM image acquisition.

\section{List of captions}

FIG 1. \textbf{(a)} Zoomed out and tilted SEM image of a pair of nanobeam devices. The metal contact pads are seen away from the cavity region and the diagram indicates approximately where electrical probes are placed. The N-type doping is seen as darker grey and the P-type doping is outlined in white dashed lines. \textbf{(b)} Zoomed in SEM of the yellow box region in \textbf{(a)}. The beam is deflected down by a small amount likely due to strain from the GaAs/AlGaAs interface. \textbf{(c)} Top view SEM of a nanobeam cavity. The scale bar is 1 $\mu$m. \textbf{(d)} FDTD calculated cavity mode electric field magnitude for the designed structure.   
\\
\\
FIG 2. \textbf{(a)} Current-voltage plots of nanobeam structures for a 400 nm intrinsic region with holes (green), 400 nm intrinsic region without holes (blue), and 5 $\mu$m intrinsic region with holes (red). Experimental results are given by the solid lines and the simulation results are shown with symbols. \textbf{(b)} Simulated steady-state injected carrier density of carriers for a lengthwise cross section slice 150 nm from the central cavity axis for a diode bias of 1.2 V. \textbf{(c)} 2D map of the hole carrier density (in cm$^{-3}$). Non-radiative recombination was modeled as acceptor-type traps at surfaces and hence there is slight hole accumulation at the edges. \textbf{(d)} 2D map of the electron carrier density (in cm$^{-3}$). \textbf{(e)} 2D map of the total current density (in A/cm$^2$).
\\
\\
FIG 3. \textbf{(a)} EL spectrum for a nanobeam device at a forward bias of 5 $\mu$A. The cavity fundamental mode is the sharp peak at 1,255 nm and the background QD emission is the broad spectrum below. Inset shows a zoom-in of the cavity peak along with a Lorentzian fit, giving a Q-factor of 2,900. \textbf{(b)} EL spectrum of the cavity used in power dependent measurements. \textbf{(c)} IR camera picture of the nanobeam cavity emission. An outline of the cavity is seen by the yellow lines and the scale bar is 5 $\mu$m. The cavity emission is bright at the center as expected and there is slight EL scattered out at the nanobeam edges. \textbf{(d)} Plot of the cavity output power and Q-factor versus injection current.
\\
\\
FIG 4. \textbf{(a)} Plot of the cavity peak wavelength versus injection current. \textbf{(b)} 2D map of the change in temperature (in $\Delta$K) of the nanobeam device for a bias voltage of 1.2 V. The cavity center only heats by about 3.3K.

\newpage

\begin{figure}[htp]
\centering
\includegraphics{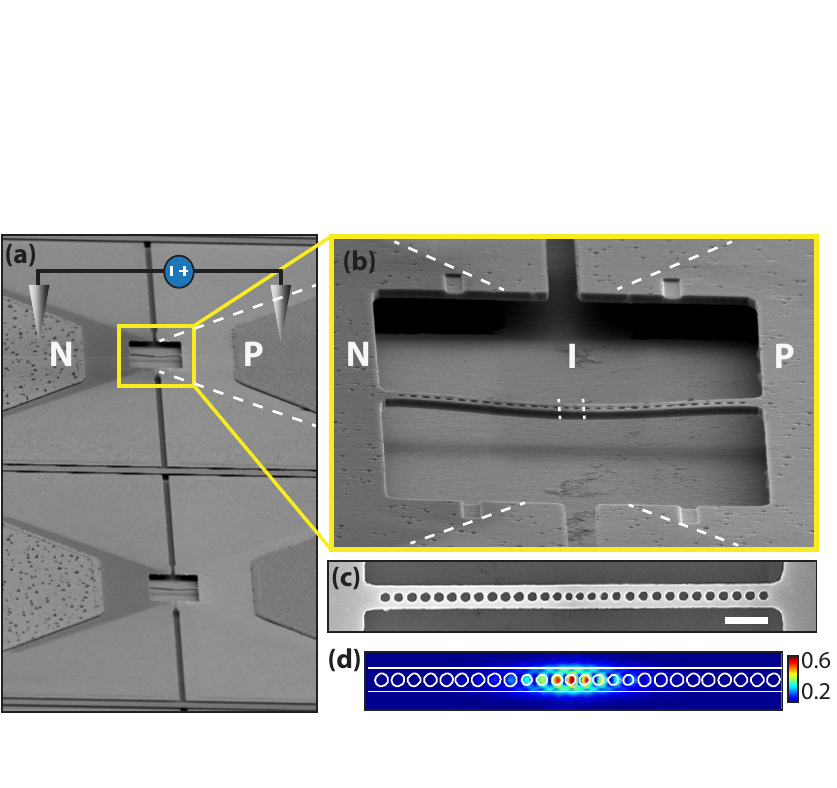}
\end{figure}

\newpage

\begin{figure}[htp]
\centering
\includegraphics{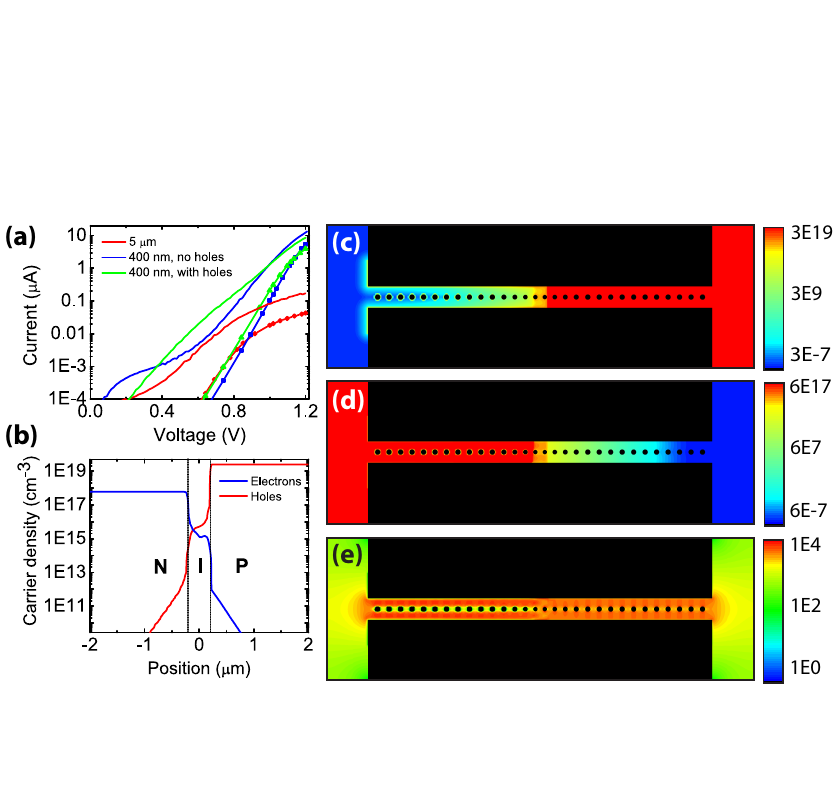}
\end{figure}

\newpage

\begin{figure}[htp]
\centering
\includegraphics{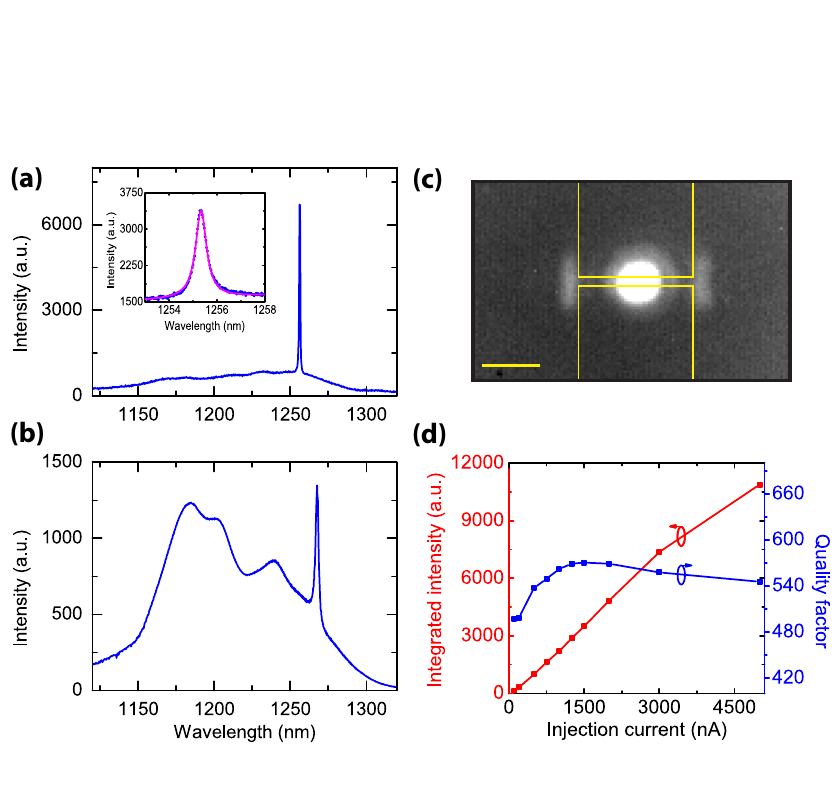}
\end{figure}

\newpage

\begin{figure}[htp]
\centering
\includegraphics{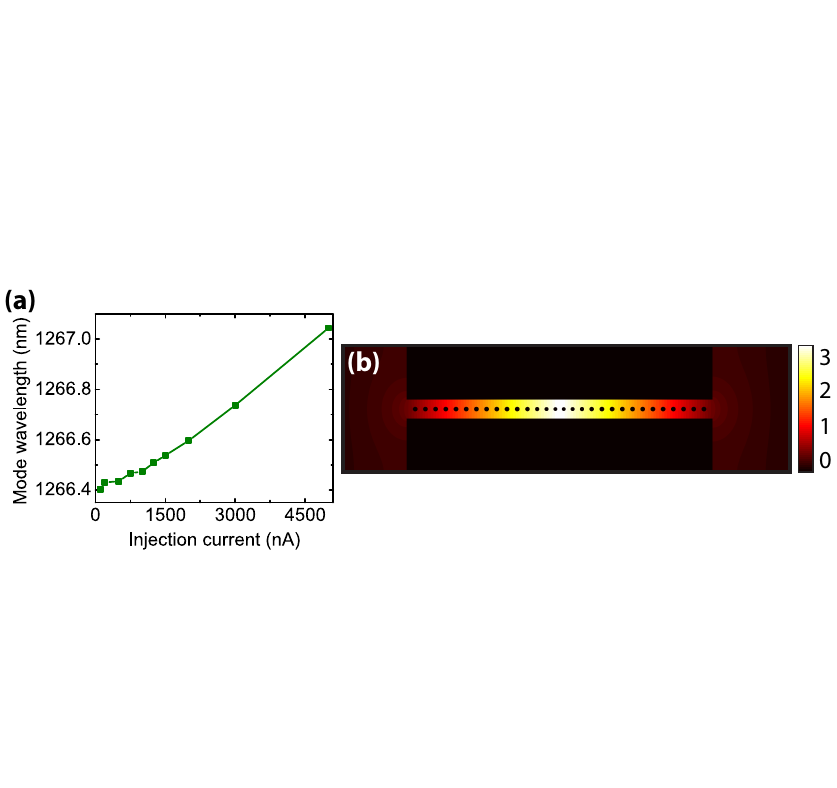}
\end{figure}

\end{document}